\documentclass[sn-mathphys,Numbered,iicol]{sn-jnl}%
\usepackage{graphicx}%
\usepackage{multirow}%
\usepackage{amsmath,amssymb,amsfonts}%
\usepackage{amsthm}%
\usepackage{mathrsfs}%
\usepackage[figuresright]{rotating}%
\usepackage[title]{appendix}%
\usepackage{xcolor}%
\usepackage{textcomp}%
\usepackage{manyfoot}%
\usepackage{booktabs}%
\usepackage{algorithm}%
\usepackage{algorithmicx}%
\usepackage{algpseudocode}%
\usepackage{program}%
\usepackage{listings}%
\providecommand{\dif}[0]{\mathrm{d}}

\providecommand{\wfac}[0]{a_w}
\providecommand{\trh}[0]{T_{\rm rh}}
\providecommand{\mpl}[0]{m_{\rm Pl}}
\providecommand{\ybap}[0]{Y_B^{\rm ap}}
\providecommand{\trhepsplane}[0]{\log (\trh/M) - \log \epsilon}
\begin{document}

\title[ ]{Washout processes in post-sphaleron baryogenesis from way-out-of-equilibrium decays}
\author*[1,2]{\fnm{J.} \sur{Racker}}%
\affil*[1]{\orgdiv{Instituto de Astronom\'{\i}a Te\'orica y Experimental (IATE)}, \orgname{Universidad Nacional
de C\'ordoba (UNC)~- Consejo Nacional de Investigaciones Cient\'{\i}ficas y T\'ecnicas
(CONICET)}, \orgaddress{\street{Laprida 854}, \city{C\'ordoba}, \postcode{X5000BGR}, \state{C\'ordoba}, \country{Argentina}}}

\affil*[2]{\orgdiv{Observatorio Astron\'omico de C\'ordoba (OAC)}, \orgname{Universidad Nacional
de C\'ordoba (UNC)}, \orgaddress{\street{Laprida 854}, \city{C\'ordoba}, \postcode{X5000BGR}, \state{C\'ordoba}, \country{Argentina}}}

\abstract{We study washout processes in post-sphaleron baryogenesis, a mechanism where the matter-antimatter asymmetry is generated in the decay of exotic particles after the electroweak phase transition. In particular we focus, in a quite model independent way, on those scattering processes that have an amplitude proportional to the CP asymmetry. We find that when the scatterings involve only massless particles, the washouts are very severe for light decaying particles (with masses below a few hundred of GeV) and successful baryogenesis is only possible in a small portion of parameter space. Instead, if even a very light particle participates in these processes, the allowed region of parameter space opens considerably, although the final amount of baryon asymmetry may differ significantly from the expression which is typically used and neglects washouts. Furthermore, we analyze washouts from the non-thermal spectrum of energetic particles produced in cascade decays and indicate in which models they can be relevant. 
}

\maketitle

\section{Introduction}\label{sec1}
Baryogenesis via leptogenesis and electroweak baryogenesis have been widely studied as mechanisms to explain the matter-antimatter asymmetry of the universe. In particular, leptogenesis is motivated by the possibility to explain within the same simple model the baryon asymmetry and the smallness of neutrino masses. However, the origin of the cosmic asymmetry remains a completely open problem and it is worth considering other mechanisms, specially if they involve different physics and testable consequences.

An intriguing possibility in this direction is post-sphaleron baryogenesis (name coined in~\cite{Babu2006}, see also previous works~\cite{Dimopoulos:1987rk,Cline:1990bw,Mollerach:1991mu,Benakli:1998ur}), where the baryon asymmetry is produced in the CP and baryon number violating decays of an exotic particle after the electroweak phase transition, and therefore not relying in electroweak sphalerons as the source of baryon number violation. This mechanism has the desirable feature of being accessible to experimental exploration and compatible with cosmological scenarios which require low reheating temperatures (see also~\cite{Davidson:2000dw}). 

The mechanism has been implemented in different models~\cite{Babu:2006wz,Kohri:2009ka,Allahverdi:2010rh,Gu:2011ff,Allahverdi:2010im,Allahverdi:2013mza,Allahverdi:2013tca,Cheung:2013hza} and observational consequences analyzed in several works~\cite{Babu:2008rq,Babu2013,An:2013axa,Cui:2014twa,Patra2014,Allahverdi:2017edd,Bell2018}, see also~\cite{Barrow2022} for a review,~\cite{Baldes:2014rda} for the role of CP violation in scatterings and~\cite{Ghalsasi:2015mxa,Aitken:2017wie,Elor:2018twp,Asaka:2019ocw} for another kind of very low scale baryogenesis mechanism. The final amount of baryon asymmetry, normalized to the entropy density, is typically estimated as $Y_B^f \simeq \epsilon \tfrac{\trh}{M}$, a relation derived 
neglecting washout processes 
e.g. in~\cite{kolb90}, where $\epsilon$ is the mean amount of baryon asymmetry per decay of the exotic particle with mass $M$, and $\trh$ is the reheating temperature. 
In this work we want to analyze, in a quite model independent way, the importance of some washout processes, particularly those scatterings that have an amplitude proportional to the CP asymmetry, and assess the accuracy of the expression given before. In some models this kind of washout has been included~\cite{Arcadi:2015ffa,Grojean:2018fus,Pierce:2019ozl} (see also the discussion in~\cite{Cui:2012jh}), but the effects from the universe becoming matter dominated were taken into account just by a global entropy dilution correction factor. We solve a set of Boltzmann equations (BEs) where the contribution of the massive decaying particle to the energy density of the universe is included and compare with previous works. Moreover, we also consider for the first time (to the best of our knowledge), washouts from the non-thermal spectrum of energetic particles generated from the decays of the massive long-lived particles.

This work is organized as follows: In Sect.~\ref{sec:thermal} we identify the main parameters that determine the baryon asymmetry in order to perform an analysis as model independent as possible, then we write an appropriate set of BEs and study  washouts from particles with a thermal distribution. Next, in Sect.~\ref{sec:non-thermal}, we analyze non-thermal washouts and finally, we conclude in Sect.~\ref{sec:conclusions}.

%%%%%%%%%%%%%%%%%%%%%%%%%%%%%%%%%%%%%%%%%%%%%%%%%%%%%%%%%%%%%%%%%%%
%%%%%%%%%%%%%%%%%%%%%%%%%%%%%%%%%%%%%%%%%%%%%%%%%%%%%%%%%%%%%%%%%%%%%
\section{Washouts from thermal processes}
\label{sec:thermal}
We want to analyze washout processes in scenarios where the baryon asymmetry is generated  in the decay of particles with very long lifetimes, 
so that they contribute significantly to the energy density of the universe during the decay epoch. This can be done, at least partially, in a model independent way by identifying the main parameters for baryogenesis, which are: the mass $M$ of the decaying particle that will be called $X$, its lifetime given by the inverse of the decay width $\Gamma$, the mean amount of baryon asymmetry generated per decay, $\epsilon$, and finally the highest among the masses of the decay products involved in washout processes, to be denoted by $m$.
For large enough lifetimes, the energy density of the universe is dominated by the $X$-field prior to decays, so that baryogenesis starts in a matter-dominated epoch. The temperature of the universe will evolve differently compared to a radiation dominated universe and we want to take this into account in our study of washout effects. 
Instead, we will take the number of relativistic degrees of freedom to be constant, 
although their change with temperature may have an important effect when decays occur near Big Bang nucleosynthesis.  An appropriate set of BEs for this scenario is therefore~\cite{kolb90}
\begin{align}
\label{eq:be}
\dot{T} & = - H T +\frac{15}{2 \pi^2 g_*} \frac{1}{T^3} \Gamma M n, \nonumber \\
\dot{n} & = -3 H n - \Gamma n, \\
\dot{n_B} & = -3 H n_B
 + \epsilon \Gamma n - \Gamma_{\rm wo} n_B, \nonumber
 \end{align}
where $T$ is the temperature, $n$ is the number density of $X$, $n_B$ is the baryon density asymmetry, $g_*$ is the number of relativistic degrees of freedom, taken to be constant and equal to 106.75 (the Standard Model value), and $\Gamma_{\rm wo}$ represents the washout rate. Here derivatives are with respect to time and the Hubble rate, $H$, is given by
\begin{align}
    H^2=\left(\frac{\dot{R}}{R}\right)^2=\frac{8 \pi}{3} \frac{1}{m_{\rm Pl}^2} \left( M n + \frac{\pi^2}{30} g_* T^4 \right).
\end{align}
Choosing the scale factor $R$ instead of time as the independent variable, we get the BEs to be integrated numerically in our analysis:
\begin{align}
\label{eq:be2}
T' & = -\frac{T}{R} + \frac{15}{2 \pi^2 g_*} \frac{1}{T^3} \frac{\Gamma}{RH} M n, \nonumber \\
n' & = -\frac{3n}{R} - \frac{\Gamma}{RH} n, \\
n_B' & = -\frac{3n_B}{R}
 + \epsilon \frac{\Gamma}{RH} n - \frac{\Gamma_{\rm wo}}{RH} n_B. \nonumber
 \end{align}
 In the baryogenesis mechanism we are considering, the CP even phase necessary to have CP violation comes from on-shell contributions to one-loop diagrams. This implies the existence of baryon number violating processes with an amplitude proportional to the size of the CP asymmetry, $\epsilon$, and a corresponding rate proportional to $\epsilon^2$. These processes can washout the baryon asymmetry very efficiently at low temperatures, so that typically there is a lower bound for the
 scale of thermal baryogenesis around $M \sim 10^5$~GeV, unless washout effects are reduced via some or a combinations of ways, which have been analyzed in detail e.g. in~\cite{Racker:2013lua}. One possibility is precisely the one we are interested in this work, namely to have a kind of particle that could be somehow produced in the primitive universe, but has a very small decay width  and therefore decays when $T \ll M$.
 In addition, we will see that for low values of $M$, it is also necessary that some of the decay products have non-negligible masses so that the washout processes become Boltzmann suppressed. It should be noted that, unlike the scenario analyzed in~\cite{Racker:2013lua} for baryogenesis above the electroweak phase transition, this requirement is naturally met by involving quarks of the second and third generations (with exotic particles being also an option in some models, e.g.~\cite{Arcadi:2015ffa,Pierce:2019ozl}). 
 
 On the contrary, we expect washouts form inverse decays to be largely irrelevant, because they are proportional to $\Gamma/H(T=M)$, which is precisely very small for long-lived particles and, moreover, their effect does not depend on the scale of baryogenesis (see e.g.~\cite{Racker:2013lua}). Therefore we include in $\Gamma_{\rm wo}$ only the processes mentioned above and choose as a fiducial rate the one used in~\cite{Racker:2013lua} for the inert doublet model:
 \begin{align}
 \epsilon = & \, h \frac{3}{16 \pi} \\
    \Gamma_{\rm wo} \, n^{\rm eq}(T) = & \, \wfac \, h^2 \frac{1}{32 \pi^5} \frac{T^6}{M^2} \nonumber \\ &\, \times \int_{x_m}^\infty \dif x \frac{\left(x^2-x_m^2\right)^4}{x^4} K_1(x) ,
    \label{eq:thermalwashoutrate}
 \end{align}
where $n^{\rm eq}$ is the equilibrium number density of a massless particle with two degrees of freedom, $x_m = m/T$, $K_1(x)$ is a modified Bessel function of the second kind, and $h$ parametrizes the size of both,  $\epsilon$ and $\Gamma_{\rm wo}$, as explained above (it includes couplings and a ratio of masses). Recall that $m$ is the highest among the masses of the external particles involved in the washout processes, so that the rate $\Gamma_{\rm wo}$ becomes Boltzmann suppressed for $T < m$. Finally, the numerical constant $a_w$ parametrizes, in a global way, how the relation between the sizes of $\epsilon$ and $\Gamma_{\rm wo}$ depend on the particular model under consideration. For most calculations we take $a_w=1$, but we will also comment on how results depend on its value (for a related analysis see~\cite{Racker:2013lua}). 

\begin{figure*}[!t]%
\centering
\includegraphics[width=1.0\textwidth]{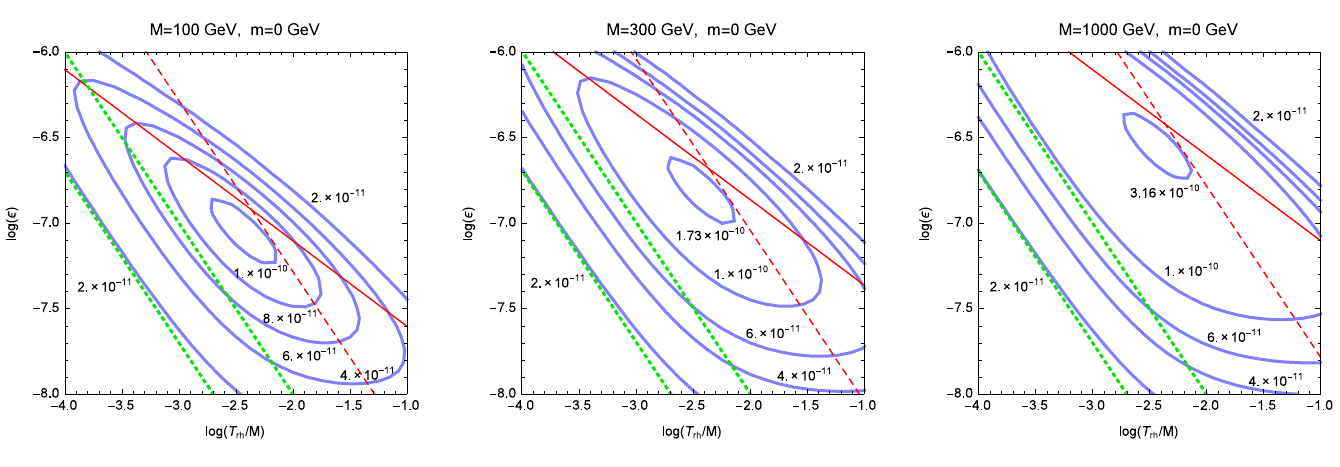}
\caption{Level curves (solid blue lines) of the baryon asymmetry in the $\trhepsplane$ plane, for $m=0$ and $M=100, 300$, and $1000$~GeV. The two green dotted lines give the isolines of $\ybap=2\times 10^{-11}$ and $10^{-10}$. The solid and dashed red lines indicate where $\Gamma_{\rm wo} = H$ at $t \simeq \tau$, with two different criteria explained in the text.} 
 \label{fig:fig1}
\end{figure*}
Next we show and analyze the results of integrating the BEs for different values of the parameters. We start the integration at $R=R_i$ (whose value is arbitrary) and take as initial conditions $T (R_i) \equiv T_i=10\,M$, $n(R_i)=n^{\rm eq}(T_i)$, and $n_B(R_i)=0$. In Fig.~\ref{fig:fig1} we plot level curves of the final baryon asymmetry normalized to the entropy density, $Y_B^f \equiv n_B^f/s$ (with the observed value being $Y_B \simeq 8.65\times 10^{-11}$), in the $\trhepsplane$ plane, for $m=0$ (massless decay products) and $M=100, 300$, and $1000$~GeV. We have chosen as independent variables $\epsilon$ (which is directly proportional to the parameter $h$) and $\trh/M$ (with the reheating temperature, $\trh$, proportional to $\sqrt{\Gamma}$), in order to ease the comparison with a simple and commonly used expression for $Y_B^f$, which is accurate for $\trh/M \ll 1$ in the absence of washouts (see e.g.~\cite{kolb90}), namely
\begin{equation}
    Y_B^f \approx \ybap \equiv \epsilon \; \frac{\trh}{M}.   
    \label{eq:aprox}
\end{equation}
The reheating temperature is defined as the temperature of the thermal bath at $t \simeq \tau$, with $\tau=\Gamma^{-1}$ the lifetime of $X$ and is given by~\cite{kolb90}:
\begin{equation}
\label{eq:tempreheating}
    \trh \simeq 0.55 \, g_*^{-1/4} \, (\mpl/\tau)^{1/2}.
\end{equation}
Therefore the isolines of constant $\ybap$ are just diagonal lines with slope -1 in the $\trhepsplane$ plane (some of these lines are explicitly drawn in the plots). 

As a second reference for comparisons, we have drawn two red lines (solid and dashed), with each line marking the boundary between the regions with $\Gamma_{\rm wo} > H$ (above the line) and $\Gamma_{\rm wo} < H$ (below the line) at the crucial time $t \simeq \tau$, for two different interpretations of these statements, which we explain next. Taking the washout rate from Eq.~\eqref{eq:thermalwashoutrate} with $m=0$ and $a_w=1$, one obtains, after performing the integral and expressing $h$ in terms of $\epsilon$,
\begin{equation}
  \frac{\Gamma_{\rm wo}}{H}\Bigr\rvert_{t=\tau} = \epsilon^2 \, \frac{2^6}{3^2 \pi} \, \frac{1}{M^2} \, T^3 \, \frac{1}{H(\tau)}. 
 \end{equation}
The Hubble rate at $t=\tau$ is related to $\trh$ by~\cite{kolb90}
\begin{equation*}
    H(\tau) \simeq \left(\frac{8 \pi^3 g_*}{90}\right)^{1/2} \frac{\trh^2}{\mpl}, 
\end{equation*}
 therefore, imposing the condition $\Gamma_{\rm wo} = H$ at $t=\tau$, we get the relation
 \begin{equation}
  \epsilon^2 \simeq \frac{3 \, \pi^{5/2}}{2^5 \, \sqrt{5}} \, \sqrt{g_*} \frac{M}{\mpl} \, \frac{M}{\trh} \left(\frac{\trh}{T}\right)^3 \, .
 \end{equation}
 Finally we want to specify a value of $T$ in this expression (coming from the dependence of the washout rate on the temperature). Choosing $T=\trh$ yields the condition described by the red solid line in the plots. But another criteria for comparisons can be to choose $T=T_D$, with $T_D$ representing the temperature just prior to decays (in an instantaneous decay approximation), which is given by~\cite{kolb90}
 \begin{equation}
     \left(\frac{\trh}{T_D}\right)^3 = \frac{S_{\rm after}}{S_{\rm before}} \simeq 1.83 g_*^{1/4} \frac{M Y_i \tau^{1/2}}{\mpl^{1/2}},    
 \end{equation}
 where $S_{\rm after}/S_{\rm before}$ is the ratio  between the entropy after and before decays, and $Y_i=n(T_i)/s(T_i)$. Expressing $\tau$ in terms of $\trh$ one arrives at another relation between $\epsilon$ and $\trh/M$, which we have represented with the dashed red line in the figures.

 Independently of whether $\trh$ or $T_D$ is used for establishing the condition $\Gamma_{\rm wo}=H$ at $t=\tau$, it is apparent from the plots that the baryon asymmetry $Y_B^f$ starts to deviate from the approximate value $\ybap$ at regions substantially below either of the red lines. Therefore the condition $\Gamma_{\rm wo}<H$ at $t \simeq \tau$ for neglecting washouts should be used with care.

 The three plots in Fig.~\ref{fig:fig1} also illustrate the scaling with $M^{1/2}$ of $Y_B^{f}$ close to its maximum, which was discussed in~\cite{Racker:2013lua}. Notice e.g. the similarities, apart from a shift in parameter space,  of the level curves $1 \times 10^{-10}$, $\sqrt{3} \times 10^{-10}$, and $\sqrt{10} \times 10^{-10}$ for $M=100, 300$, and $1000$~GeV, respectively. 

\begin{figure*}[!t]%
\centering
\includegraphics[width=0.85\textwidth]{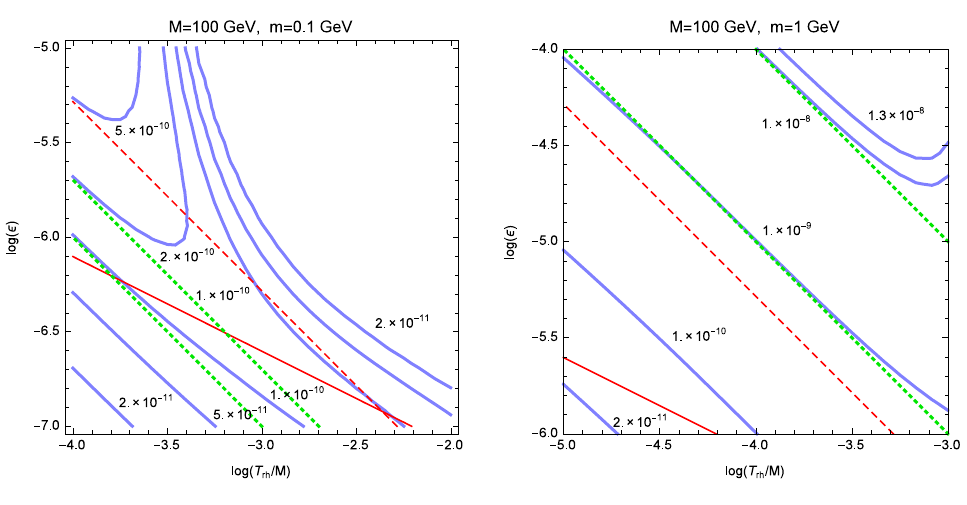}
\caption{Similar to Fig.~\ref{fig:fig1}, but for $M=100$~GeV and $m=0.1$ ($1$~GeV) in the left (right) plot. Moreover, here the two green dotted lines give the isolines of $\ybap=10^{-10}$ and $2\times 10^{-10}$ (left plot), and $\ybap=10^{-9}$ and $10^{-8}$ (right plot). The red curves have been determined as in Fig.~\ref{fig:fig1}, with $m=0$, therefore they lose meaning for $\trh < m$.} 
 \label{fig:fig2}
\end{figure*}

For relatively light decaying particles, $M \sim$ few 100~GeV, the washouts we are considering, which are proportional in magnitude to the square of the CP asymmetry, are very severe and preclude successful baryogenesis except in a small region of parameter space. However, if some of the external particles that participate in these washout processes are massive, washout effects are drastically reduced. This is illustrated in Fig.~\ref{fig:fig2}, for $M=100$~GeV, where we have taken $m=0.1$ and $1$~GeV. It is apparent that when $\trh$ becomes somewhat smaller than $m$, large values of the baryon asymmetry can be obtained by increasing $\epsilon$ without washouts becoming a problem, even if $m$ is very small relative to $M$ (left plot). Still, in the left plot the differences between $Y_B$ and the approximate value $\ybap$ are apparent in most regions of the parameter space. Instead, in the right plot with $m=1$~GeV, we have covered a region of parameter space with lower values of $\trh$, where $Y_B$ can be very large and the approximation in Eq.~\eqref{eq:aprox} is really accurate. Note that we have still drawn the red curves, but they are determined as above, with $m=0$, therefore they lose meaning for $\trh < m$. 

Let us summarize some other issues we have analyzed:
\begin{itemize}
\item In the plots we have taken the number of relativistic degrees of freedom constant and equal to $g_*=106.75$. However this is a bad approximation for $T \lesssim 100$~MeV. To get an idea of how much $Y_B^f$ changes for lower values of $g_*$, we have repeated some of the scans with $g_*=20$ and found that the baryon asymmetry in relevant regions of the parameter space (i.e. where the baryon asymmetry does not differ much from its observed value) gets reduced by a factor of around 3 (compared to the one obtained with $g_*=106.75$). Therefore this is a quite important effect to keep in mind, but a more precise study should use a modified set of BEs (the ones we have solved cease to be valid when $g_*$ varies with the temperature). 
\item The kind of washout we are analyzing in this section has been included in some works concerning particular models, however the set of BEs used correspond to a universe that is all the time dominated by radiation. After integrating the BEs, the final baryon asymmetry is corrected by the dilution factor $S_{\rm after}/S_{\rm before}$ mentioned above. We have repeated some of the scans using this approximate approach (to make plots like those in Figs.~\ref{fig:fig1} and~\ref{fig:fig2}) and found that the resultant baryon asymmetry can be overestimated by a factor around 2 (again in some relevant regions of parameter space).
\item Much of model dependencies can be captured by the parameter $a_w$ that allows to adjust the relation between the sizes of the CP asymmetry and washout rates. E.g. we expect it to get larger as the number of decay products increases (in the CP- and B-violating decays of $X$). In the plots we have taken $a_w=1$, but note that a change in $a_w$ can be absorbed by a redefinition of $h^2$ with the corresponding change in $\epsilon$ being $1/\sqrt{a_w}$. Since $Y_B^f$ is directly proportional to $\epsilon$, we can see that compared to our plots, an increase in $a_w$ will shift level curves in the plane while reducing $Y_B^f$ by a factor $\sqrt{a_w}$. Therefore model dependencies are quite mild in this regard. 
\item A difference that can be more significant among models is how $X$ is produced in the first place and whether or not it may decay or annihilate via baryon number conserving processes. All this would lead to a different value of $n(T_i)$ and a corresponding change in $Y_B^f$, which depends linearly on $n(T_i)$.
\end{itemize}

We have seen that when massive particles participate in the washout processes, they become Boltzmann suppressed when $\trh \lesssim m$. This is true for particles with a thermal distribution, however there is also a non-thermal spectrum of energetic particles generated from the decays of $X$ that may induce washouts. Therefore in the following section we analyze this effect. 
 
\section{Washouts from non-thermal processes}
\label{sec:non-thermal}
In post-sphaleron baryogenesis the cosmic asymmetry is generated in way-out-of-equilibrium conditions.  A particle with a mass much larger than the temperature $T$ of the thermal bath, can decay into particles which may also be quite massive relative to $T$, and therefore the related washout processes, which require the presence (or creation) of these massive particles in the thermal bath, are almost banned. This is a quite extreme situation of energetic processes occurring in only one direction. It is reasonable to ask if the only ``price to pay" for this is the large amount of entropy generation. Indeed, a more detailed view reveals that there are actually three main components in this scenario: the population of $X$ particles which dominates the energy density of the universe, the thermal bath made of relativistic particles in thermal equilibrium at a temperature $T$, and a non-thermal spectrum of particles originated from the decays of $X$. The decay products of $X$ are very energetic, and although it is to expect that they will thermalize very fast, some amount of non-thermalized, high-energy particles, will be present during the whole decay epoch (when the baryon asymmetry is generated). These energetic particles may induce washout processes which are not possible for the low-energy components of the thermal bath. Next we proceed to estimate the size of this effect.
Our discussion will be based on~\cite{Drees:2021lbm}, which includes an analytical approximation for the non-thermal spectrum coming from the decay products of a heavy particle. The approximation takes into account the Landau-Pomeranchuk-Migdal (LPM) effect~\cite{Landau:1953um,Migdal:1956tc}, which slows
down thermalization and therefore leads to a density of non-thermal particles that is higher than the value expected from more naive estimations.

For concreteness it is assumed that the particles $X$ decay into two  ultra-relativistic particles which in turn generate a non-thermal spectrum of daughter particles, to be denoted by $\tilde n(E)$, and defined so that 
$\int_T^M \tilde n(E)\, \dif E = n_{\rm nt}$, with $n_{\rm nt}$ the total number density of non-thermalized particles. It is convenient to write
\begin{equation}
    \tilde n(E) \equiv \tilde N_M \, \bar n(E),
\end{equation}
where $\tilde N_M$ is a normalization factor given by 
\begin{equation}
    \tilde N_M = \frac{2 n \Gamma}{\Gamma^{\rm split}_{\rm LPM} (M/2)} \simeq \frac{2 n \Gamma}{2 \alpha^2 \sqrt{\tilde g_*} T},
\end{equation}
$\alpha$ is the coupling strength of the gauge interaction responsible for thermalization, and $\tilde g_*$ is the number of degrees of freedom in the thermal bath that couple to this gauge interaction. The last equality depends on the value chosen for an infrared cutoff, which is one of the uncertainties for the correct determination of the non-thermal spectrum (see~\cite{Drees:2021lbm}).
After changing variables from $E$ to $x=E/T$, so that $\bar n(E) = \bar n(x)/T$, the following analytical approximation for $\bar n(x)$ was obtained in~\cite{Drees:2021lbm}:
\begin{align}
    &\bar{n}(x,x_M)=\delta(x-x_M) \, +\\
    &\frac{\left[a\left(\frac{x}{x_M}\right)^{-3/2}\left(1-\frac{x}{x_M}\right)^{-b}+c\right]\left(1-\frac{2}{\sqrt{x_M}}\right)}{\sqrt{x_M}\left(1-\sqrt{2/x}\right)^{5/4}}, \nonumber
\end{align}
with $x_M=\tfrac{M}{2 T}$, and $a, b, c$ are three parameters whose values are set in order to get a good fit to numerical calculations. As in~\cite{Drees:2021lbm}, we take them equal to $a = 0.39, b = 0.48, c = 0.37$.

The flux of non-thermal energetic particles induces a washout rate, $\Gamma_{\rm wo}^{\rm nt}$, which is given approximately by
\begin{equation}
\label{eq:nonthermalrate}
    \Gamma_{\rm wo}^{\rm nt} = \sigma \int_{m}^{M/2} \tilde n(E) \dif E,
\end{equation}
where the cross section $\sigma$ has been considered as roughly constant (because of the high energies involved). We take as a reference the cross section from~\cite{Racker:2013lua}, 
\begin{equation}
\label{eq:crosssection}
    \sigma = \frac{1}{ 2 \pi M^2} h^2.
\end{equation}
Note that in Eq.~\eqref{eq:nonthermalrate} we start the integration at $E=m$, i.e. we only consider the flux of non-thermal particles with energies larger than the masses of the external particles in the baryon-number-violating processes, so that the cross section can indeed be approximated to a constant. 

 The washout term from non-thermal processes can be obtained gathering the previous expressions and defining for convenience $Y=n/s$, with $s=2 \pi^2 g_* T^3/45$ the entropy density:
 \begin{equation}
 \label{eq:nonthermalwashout}
     \frac{\Gamma_{\rm wo}^{\rm nt}}{H}  =  \frac{2^8 \, \pi^3}{3^4\, 5} \, \epsilon^2 \, \frac{g_*}{\alpha^2 \sqrt{\tilde g_*}} \,  Y \, \frac{\Gamma}{H}\, \frac{T^2}{M^2}  \int_{m}^{M/2} \bar n(E) \, \dif E. 
 \end{equation}
The ratio $\Gamma_{\rm wo}^{\rm nt}/H$ should be order 1 or larger at $t\sim \Gamma^{-1}$, i.e. at $T \sim T_{\rm rh}$, for non-thermal washouts to be important. Taking as a very crude approximation $\int_{m}^{M/2} \bar n(E) \, \dif E \sim M/m$ (as if the energy $M$ of the parent particle were distributed among $M/m$ daughter particles with mass $m$, see~\cite{Drees:2021lbm}), we get, very roughly, \begin{equation}
\label{eq:nonthermalaprox}
  \frac{\Gamma_{\rm wo}^{\rm nt}}{H}\Bigr\rvert_{T=T_{\rm rh}}  \sim \, \epsilon^2 \, \frac{10}{\alpha^2 \sqrt{\tilde g_*}} \, \frac{T_{\rm rh}^2}{M^2} \, \frac{M}{m}. 
 \end{equation}

\begin{figure*}[!t]%
\centering
\includegraphics[width=0.5\textwidth]{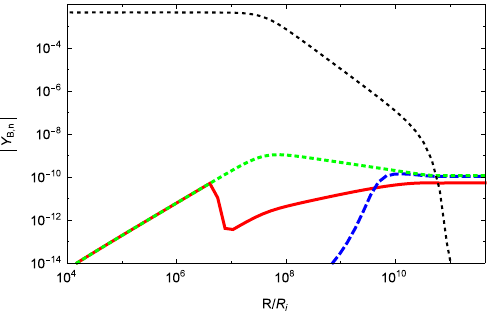}
\caption{Absolute value of $Y_n\equiv n/s$ (dotted black curve), $Y_B$ without including washouts (dotted green curve), including only washouts from thermal processes (dashed blue curve), and including only washouts from non-thermal processes (solid red curve), as a function of the scale factor $R$ normalized to some arbitrary initial value $R_i$. We have taken $M=10^6$~GeV, $M_N=100$~GeV, $m=1$~GeV, $\epsilon=0.05$, and $\Gamma/H(T=M)=10^{-15}$, so that $\trh \approx 22$~MeV. Furthermore, we have set $\alpha^2 \sqrt{\tilde g_*} = 1$. The final baryon asymmetry $Y_B^f \simeq 10^{-10}$ when non-thermal washouts are ignored, while $Y_B^f \simeq 5 \times 10^{-11}$ if they are included.} 
\label{fig:fignt}
\end{figure*}

 The interesting case is when $m \gtrsim T_{\rm rh}$ (for $m \ll T_{\rm rh}$ washouts from thermal processes are clearly dominant), therefore even for a large CP asymmetry, the ratio in Eq.~\eqref{eq:nonthermalaprox} cannot be much larger than $\trh/M \ll 1$. The conclusion seems to be that non-thermal washouts are negligible, but the previous analysis requires a modification for certain types of models like the one proposed in~\cite{Allahverdi:2010rh}. When the baryon asymmetry is not produced directly in the decays of $X$, but in the decays of a lighter particle $N$ which is one of the decay products of $X$, the above estimation changes to  
\begin{equation}
\label{eq:nonthermal2}
  \frac{\Gamma_{\rm wo}^{\rm nt}}{H}\Bigr\rvert_{T=T_{\rm rh}}  \sim \, \epsilon^2 \, \frac{10}{\alpha^2 \sqrt{\tilde g_*}} \, \frac{T_{\rm rh}^2}{M_N^2} \, \frac{M}{m}, 
 \end{equation}
where $M_N$ is the mass of $N$, which can be much smaller than $M$. This is because $M_N$ must be used instead of $M$ in Eq.~\eqref{eq:crosssection} (note that this may be a conservative estimation of the cross section of the washout processes, because the mass of the mediator might be similar to $M_N$, but we are not considering, neither possible resonant enhancements, nor an increase of $\sigma$ with the energy of the non-thermal particles). We find that for certain values of the parameters, the non-thermal washouts can indeed be relevant in this class of model. For the purpose of illustration we take $M=10^6$~GeV, $M_N=100$~GeV, $m=1$~GeV and $\trh \approx 20$~MeV, and plot in Fig.~\ref{fig:fignt} the evolution of the baryon asymmetry as a function of the scale factor, without including washouts, including only the washouts from thermal processes, and including only the washouts from non-thermal processes. The curves have been obtained integrating the BEs from the previous section after multiplying the source term by the branching ratio of $X$-decays into $N$ (taken equal to 0.1), and using Eq.~\eqref{eq:nonthermalwashout} for the washout term from non-thermal processes (with $M_N^2$ instead of $M^2$ in the denominator). It is clear that at high temperatures, $T \gtrsim m$, the washouts from thermal processes are important and dominant, however at lower temperatures they become Boltzmann suppressed and do not affect the final value of the baryon asymmetry, which is mainly generated at late times. On the contrary, washouts from non-thermal processes are irrelevant at $T \gtrsim m$, but become important at lower temperatures and reduce the final baryon asymmetry by a factor $\sim 2$ (indeed, for physical reasons and numerical convenience we have set these washouts to zero at high temperatures in the integration of the BEs, but the final asymmetry does not depend on exactly when they are turned on).  

%%%%%%%%%%%%%%%%%%%%%%%%%%%%%%%%%%%%%%
%%%%%%%%%%%%%%%%%%%%%%%%%%%%%%%%%%%%%%%
%%%%%%%%%%%%%%%%%%%%%%%%%%%%%%%%%%%%%%

\section{Conclusions}
\label{sec:conclusions}
We have studied the importance of washouts in baryogenesis from late decays of heavy particles after the electroweak phase transition, using Boltzmann equations appropriate for a universe that becomes matter dominated previous to the decays. 
By identifying the main parameters that determine the baryon asymmetry we could do a quite model independent analysis of several issues (possible model dependencies were also discussed). 
In Figs.~\ref{fig:fig1} and~\ref{fig:fig2} we have plotted level curves of the baryon asymmetry 
in the $\trhepsplane$ plane, allowing an easy comparison with the approximation $\ybap = \epsilon \,\trh/M$ that is often used to estimate the asymmetry neglecting washouts. We have found that washouts can be important even in regions of parameter space that clearly violate the rough condition $\Gamma_{\rm wo} < H(t \simeq \tau)$ that is frequently used to disregard washouts. Moreover, we have also compared with more detailed treatments that include washouts, but they do so in BEs that assume a radiation dominated universe and correct at the end with a dilution factor due to entropy generation. We have found that this approach may overestimate the baryon asymmetry by a factor $\sim 2$ in relevant regions of parameter space, i.e. when the calculated asymmetry is around the observed value.

When washout processes involve external particles of mass $m$  and $\trh \lesssim m$ (with $\trh$ given by Eq.~\eqref{eq:tempreheating}), their rate becomes Boltzmann suppressed for particles with a thermal distribution. However, there is also a non-thermal spectrum of energetic particles generated from the decays of the long-lived particle. We have estimated the washout effects from this non-thermal component, finding that they are typically negligible, except in certain kinds of models (see Eqs.~\eqref{eq:nonthermalaprox} and~\eqref{eq:nonthermal2}).  

Another sort of washout results from many body processes where the particles with mass $m$ appear in propagators, not as external states. The rates of theses processes are expected to be small, but they are not always negligible. However, the analysis of this kind of washout is more model dependent and is left for future work. Another issue we have not addressed in detail is that for $T \lesssim 100$~MeV the number of relativistic degrees of freedom, $g_*$, decreases substantially. The BEs we have used assume that $g_*$ is constant, nevertheless, and just to make an estimation, we repeated some of the numerical scans with $g_*=20$ (instead of 106.75), finding that the baryon asymmetry in some relevant regions of parameter space gets reduced by a factor $\sim$ 3. Therefore, it could be worth considering a more detailed treatment of this effect.

\bibliography{referenciasleptogenesis3}

\end{document}